\documentclass[aps,prb,showpacs]{revtex4}
\usepackage{graphicx}
\usepackage{epstopdf}

\begin{document} 
\title{  Gauge-field rotation of 2D exciton Bose condensate in double quantum well by radial magnetic field}  
\author{ E.B. Sonin}

\affiliation{ Racah Institute of Physics, Hebrew University of
Jerusalem, Jerusalem 91904, Israel} 

\date{\today} 

\begin{abstract}
Here it is shown that a radial magnetic field leads to rotation of a Bose condensed exciton cloud due to Aharonov-Bohm effect for an electron and a hole forming an exciton. As in the case of mechanical rotation of superfluids, rotation is  accompanied by penetration of vortices into the cloud at some critical magnetic field. Penetration of vortices strongly affects the total intensity and the angular distribution of photoluminescence from the exciton cloud. This effect can be used for an effective  experimental manifestation of exciton Bose condensation.
\end{abstract} 

\pacs{71.35.Lk, 73.21.Fg, 03.65.Vf}
\maketitle

Bose condensation of excitons has already been discussed more than four decades \cite{Mos,Blatt}.  The idea to look for Bose-condensation of spatially separated electron-hole pairs (in electron-hole bilayers, in particular) is also very old \cite{KogTav,Loz,She}. The interest to this idea was revived after experimental observation of excitons in coupled doubled quantum wells, which stimulated intensive theoretical and experimental investigations \cite{But,Rap,Tim,Snoke2}. Though some experimental evidences of exciton Bose condensation in these systems were reported, the question whether Bose condensation was really observed remains unsettled \cite{Snoke}. The key problem is an unambiguous detection of phase coherence, which must exist in a Bose condensate. The present Letter suggests to probe phase coherence by studying rotation of the exciton cloud in a radial magnetic field. 

The rotation of the exciton cloud is possible due to the gauge field connected with the electromagnetic vector potential, which is responsible for the Aharonov-Bohm effect.
Originally the  Aharonov-Bohm effect was predicted for charged particles \cite{AB},  but later it was generalized on neutral particles with magnetic\cite{AC} or electric\cite{Wil,Wei} dipole momenta. 
Recently Sato and Packard \cite{Pack} reported on a plan to detect the Aharonov-Bohm effect for neutral dipoles in superfluid $^4$He.  The Hamiltonian for a (quasi)particle with an  electric dipole moment ${\mathbf d}$ is:
\begin{eqnarray}
{\cal H}={1\over 2 m} \left( {\mathbf p} -{e\over c} \tilde {\mathbf A}\right)^2={1\over 2 m} \left( {\mathbf p} -{1\over c} [ {\mathbf H}\times {\mathbf d}]\right)^2.
   \label{ham} \end{eqnarray}
 One can find the derivation of the Hamiltonian (\ref{ham}) for translational motion of a general charge aggregate in Ref.~\onlinecite{Bax}. The physical origin of the Hamiltonian (\ref{ham}) can be illustrated as the following. A rigid dipole ${\mathbf d}=e{\mathbf a}$ formed by two charges $\pm e$ connected with the position vector $\mathbf a$ feels the electromagnetic vector potential ${\mathbf A}({\mathbf r}+{\mathbf a})-{\mathbf A}({\mathbf r}+{\mathbf a}) \approx ({\mathbf a}\cdot {\mathbf \nabla }){\mathbf A}({\mathbf r} = {\mathbf \nabla }({\mathbf a}\cdot {\mathbf A}) +[ {\mathbf H}\times {\mathbf a}]$. The first term in the latter expression is a gradient of a nonsingular scalar function, which can be removed with a gauge transformation. The second term yields the ``effective''  electromagnetic vector potential $\tilde {\mathbf A} =[ {\mathbf d}\times {\mathbf H}]/e$ in the Hamiltonian (\ref{ham}). The effect of the gauge field $\tilde {\mathbf A} $ is sometimes referred as the R\"ontgen effect\cite{Bax} though the effect observed by R\"ontgen does not require the presence of the gauge field for its explanation (see below).

The velocity of a dipole is given by the standard expression ${\mathbf v} = \left({\mathbf p} -{e\over c} \tilde {\mathbf A}\right)/m$, so the gauge field $\tilde {\mathbf A}$  can produce a neutral current of dipoles.  
However, in a large cloud of uncorrelated excitons no essential currents are possible in the ground state since the gauge-field effect  will be compensated by the canonical momentum $ {\mathbf p}$. This compensation suppresses the kinetic energy. For Bose-condensed particles the situation is different. The gauge-field velocity ${\mathbf v}_d=e \tilde {\mathbf A}/mc$ is not curl-free in general, its vorticity being
\begin{eqnarray}
{\mathbf \omega} = [ {\mathbf \nabla }\times {\mathbf v }_d ]={1\over mc} [ {\mathbf \nabla }\times  [ {\mathbf H}\times {\mathbf d}] ]. 
      \end{eqnarray}
The canonical momentum ${\mathbf p} =\hbar {\mathbf \nabla}\phi$ is determined by the phase gradient, i.e., is curl-free and therefore  cannot compensate the gauge-field velocity if vortices are absent. Thus the gauge field can rotate a Bose condensate of dipoles.

\begin{figure}[!t]
\begin{center}
   \leavevmode
  \includegraphics[width=0.3\linewidth] 
  {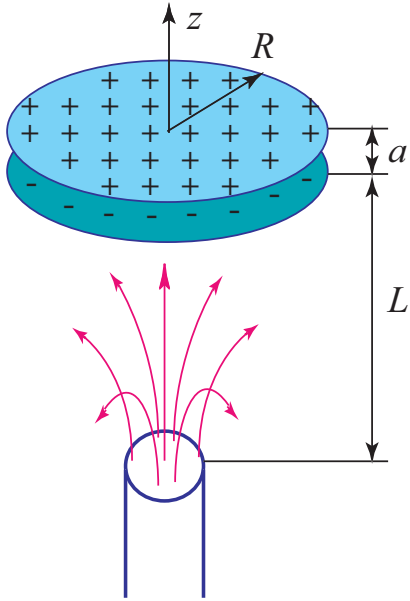}
  \caption{The exciton cloud in the electron-hole  bilayer  in the flaring magnetic field.}
 \label{fig1}
 \end{center}
\end{figure}

Excitons in a double quantum well with electrons in one well and holes in another have a component of the dipole moment $d_z=ea$ normal to the bilayer plane, where $a$ is on the order of the distance between two wells.  For an axisymmetric 2D exciton cloud  in an axisymmetric magnetic field (Fig.~\ref{fig1}) the gauge-field vorticity component along the axis $z$ is 
\begin{eqnarray}
\omega_z =  {ea\over mc}\left({\partial H_r \over \partial r}+{H_r\over r}\right),
      \end{eqnarray}
where $H_r$ is the radial in-plane component of the magnetic field. 
The radial magnetic field can be produced by a long thin solenoid (see Fig.~\ref{fig1}). The flaring magnetic field around its end is similar to that of the magnetic monopole. 
 If the distance $L$ of the exciton cloud  from the solenoid end is much larger than the cloud radius $R$ the radial magnetic field is linear with respect to $r$: $H_r=H_0 r/R$, where $H_0=H_z R/L$ is the radial magnetic field at the cloud boundary $r=R$ and $H_z$ is the axial magnetic field in the cloud center. Then the gauge field fully simulates the solid-body rotation with the angular velocity $\Omega =\omega_z/2={H_0 d/ mcR}={eH_0 a/ mcR}$. Later on we shall ignore the axial field $H_z$. Its effect can be really removed by putting a second solenoid above the cloud mirroring the first one.

The effect of rotation on a superfluid Bose liquid or gas is well studied. In general the kinetic energy of the Bose condensed exciton cloud is given by 
\begin{eqnarray}
E= {m \over 2}  \int n_2 (r) \left[{\hbar \over m}{\mathbf \nabla }\phi -{\mathbf v}_d({\mathbf r})\right]^2 d {\mathbf r},
  \label{En}    \end{eqnarray}
where $\phi $ is the phase of the Bose-condensate wave function, and integration is over the whole 2D condensate with the 2D density distribution $n_2 ({\mathbf r})$. Despite a close analogy with rotating superfluids (with the velocity ${\mathbf v}_d$ instead of the solid-body-rotation velocity ${\mathbf v}_0=[{\mathbf \Omega} \times {\mathbf r}]$) there still remains an essential difference.  The mechanically rotated superfluid is described by the energy in the rotating coordinate frame
\begin{eqnarray}
E=E_0 - {\mathbf \Omega} \cdot  {\mathbf J }= {m \over 2}  \int n_2 (r) \left[{\hbar \over m}{\mathbf \nabla }\phi({\mathbf r}) \right]^2 d {\mathbf r}
-  \hbar   \int n_2 (r){\mathbf \nabla }\phi  \cdot {\mathbf v}_0 ({\mathbf r}) d {\mathbf r}
\nonumber \\
= {m \over 2}  \int n_2 (r) \left[{\hbar \over m}{\mathbf \nabla }\phi -{\mathbf v}_0({\mathbf r})\right]^2 d {\mathbf r}-{m \over 2}  \int n_2 (r) \left[{\mathbf v}_0({\mathbf r})\right]^2 d {\mathbf r},
      \end{eqnarray}
where $\mathbf J$ is the angular momentum of the superfluid. The first term in this expression is the kinetic energy of the superfluid in the rotating coordinate frame while the second term is the centrifugal energy. The latter term is absent in the energy of the exciton cloud given by equation~(\ref{En}). 

For low magnetic fields  the phase-gradient velocity is absent, and the kinetic energy of the axisymmetric cloud of the radius $R$   is given by  
\begin{eqnarray}
E_0=\pi m  \int_0^Rn_2 (r) v_d(r)^2 r\,dr. 
      \end{eqnarray}
This energy increases with the magnetic field and at some critical magnetic field the state with a vortex becomes more stable (has a smaller energy). Assuming that  a single-quantum circulation vortex appears  at the cloud center, the energy becomes 
\begin{eqnarray}
E_v =\pi m  \int_0^Rn_2 (r) \left(v_d(r)- {\hbar  \over mr}\right)^2 r\,dr. 
      \end{eqnarray}
In an electrostatic trap for excitons with steep walls\cite{Rap} one may neglect spatial variation of the  exciton density $n_2 $.  Comparing the energies $E_v$ and $E_0$ under this assumption, one obtains that the state with a vortex becomes the ground state at the magnetic field $H_ 0$ exceeding the critical value
\begin{eqnarray}
H_{cr}={\Phi_0\over 2\pi aR}\ln{R\over r_c},
 \label{cr}     \end{eqnarray}
where $\Phi_0=hc/e$ is the magnetic-flux quantum and $r_c$ is the vortex core radius (see below),  which was used as a lower cutoff  in the integral for the vortex energy. The phase shift produced by the critical field along the closed path of the length $2\pi R$ around the cloud in the vortex-free state is $\Delta \phi = 4\pi ^2 a R H_{cr} /\Phi_0=2\pi \ln(R/r_c)$. The first vortex decreases this shift with $2\pi$. At further growth of the magnetic field more and more vortices should penetrate into the cloud eventually forming a vortex array similar to that in mechanically rotated superfluids. However, the role of vortices is opposite in the two cases. At the mechanical rotation of the container the superfluid does not rotate without vortices, but does rotate when vortices appear. At the gauge-field rotation the superfluid rotates in the vortex-free state, whereas  vortices tend to suppress rotation eventually stopping it when there is a large number of them. 

In a weakly non-ideal Bose gas the size of the core is determined by the coherence length  $\xi =h/\sqrt{m\epsilon}$, where  $\epsilon$ is the energy of boson interaction. For excitons in the double well this is  the dipole-dipole repulsion\cite{Rap} $\epsilon=4\pi e^2 an_2 /\varepsilon  $. 
Then 
\begin{eqnarray}
\xi={h\over 2e   }\sqrt{\varepsilon \over \pi m a n_2}. 
                          \end{eqnarray}
Assuming the density $n_2=10^{10}$ cm$^{-2}$, the interlayer distance $a=10^{-6}$ cm, the exciton mass $m=0.2 m_e$, where $m_e$ is the free-electron mass, and the dielectric constant $\varepsilon=10$, this yields $\xi =1.4\times 10^{-6}$ cm, which is smaller than the interparticle distance $\sim 10^{-5}$ cm. This means that the model of the non-ideal Bose gas is not not so good for the exciton gas, and the interparticle distance $n_2^{-1/2}$ is a better estimation for the core radius $r_c$ (like in superfluid $^4$He). Using it in the expression equation~(\ref{cr}) for the critical magnetic  field, one obtains for the cloud of radius $R =10~\mu$m  $H_{cr}=293$ G.

Our analysis has demonstrated that the radial magnetic field makes the exciton Bose condensate to rotate. The next question is how to detect this rotation.  Neutral currents of dipoles are  also able to produce magnetic fields though much weaker than charged currents. In fact this is the essence of original R\"ontgen's  effect\cite{R}, who detected a weak magnetic field produced by a rotating polarized dielectric disk. Referring to the electron-hole bilayer with the interlayer distance $a$ the neutral current of electron-hole pairs is a counterflow of two intralayer charged currents $\pm e j$. If the bilayer is infinite in all direction this counterflow creates a magnetic field $4\pi ej/c$ only in the interlayer space. But if the bilayer occupies a semi-infinite space and has an edge at $x=0$ there are stray fields around the edge, which are determined by the magnetic flux $4\pi eja /c$ (per unit length of the edge) exiting from the interlayer space. So the radial magnetic field at the distance $r$ from the edge is $2 eja /cr$. In fact, any inhomogeneity of the moving dipole layer must induce magnetic fields outside the layer. For example, 
there are magnetic fields induced by vortices in rotating Bose condensates of electrically polarized atoms, as was demonstrated by Leonhardt and Piwnicki\cite{LP} and Bennet {\em et al.}\cite{BBB}. These fields are quite weak. Near the edge of the cloud, where the field is maximal, it does not exceeds $\Delta H \sim 4\pi j/c =(4\pi e^2 n_2 a/m c^2)H$. Choosing for an estimation the parameter values given above, this field is equal to $\Delta H \sim 1.8\cdot10^{-7}H_0$.

Another way to detect the gauge-field effects is observation of photoluminescence. Bose-condensation of excitons in the double quantum well should lead to emission of coherent light\cite{FGO,FTM,Coh}. The dipole interaction of excitons with the electromagnetic field is linear with respect to exciton creation and annihilation operators. Their averages vanish in the Fock states with fixed numbers of excitons, but must be replaced by the Bose condensate wave function in the coherent state resulting from Bose condensation. This leads to an oscillating classical polarization component
\begin{eqnarray}
{\mathbf P}({\mathbf r})=\mbox{Re}\left\{{\mathbf d}{\psi ({\mathbf r})\over \sqrt {S}}e^{-i\omega_0t}\right\}, 
               \label{dip}           \end{eqnarray}
where $\psi ({\mathbf r})$ is the Bose-condensate wave function normalized to the total number $N$  of excitons ($\int |\psi ({\mathbf r})|^2\,d_2 {\mathbf r}=\int n_2 ({\mathbf r})\,d_2 {\mathbf r}= N$), $S$ is the area of the exciton cloud,   and $\omega_0$ is the frequency determined by the energy of the exciton 
annihilation $E_b=\hbar \omega_0$. Thus the total oscillating polarization  component $\int {\mathbf P}({\mathbf r})\,d_2 {\mathbf r}$ is proportional to $\sqrt{N}$, whereas the constant average total polarization ${\mathbf d} N$ is proportional to the total number of particles. In general the exciton dipole moment ${\mathbf d}$ has not only an axial component $d_z$ but also complex inplane  components. For example, the dipole rotating in the plane corresponds to $d_y =\pm i d_x$. A coherent light emitted by an oscillating macroscopic dipole can be treated classically\cite{Jack}. 
The light power emitted into the elementary solid body angle $d\Omega =\sin \theta d\theta d\phi$ is given by 
\begin{eqnarray}
{dW(\theta,\phi)\over d\Omega}  ={ck^4\over 8\pi}{|{\mathbf d}_\perp|^2\over S} \left|\int_S \psi({\mathbf r}) e^{i{\mathbf k}_{in} {\mathbf r}} \,d_2{\mathbf r}\right |^2,
           \label{flux}           \end{eqnarray}
where $\theta$  and $\phi$ are the polar and azimuthal angle of the wave vector ${\mathbf k}$ of light  with the inplane component ${\mathbf k}_{in}$ ($k_{in}=k\sin\theta$), and
${\mathbf d}_\perp$ is the projection of the dipole moment onto the plane normal to  ${\mathbf k}$, which determines the polarization of light. Its modulus  is given by
\begin{eqnarray}
|{\mathbf d}_\perp|^2=|d_x|^2 (1-\sin^2\theta \cos ^2\varphi)+|d_y|^2 (1-\sin^2\theta \sin ^2\varphi)
+|d_z|^2 \sin^2 \theta 
\nonumber  \\
-(d_x^*d_y+d_xd_y^*)  \sin ^2\theta \sin\varphi \cos\varphi
+(d_x^*d_z+d_xd_z^*)  \sin \theta \cos \theta   \cos\varphi
+(d_y^*d_z+d_yd_z^*)  \sin \theta\cos \theta   \sin\varphi].
                      \end{eqnarray}
Our result fully agrees with earlier calculations of the radiative decay rate using Fermi's Golden Rule\cite{And,Levitov}, if one defines the rate of emission of a photon with the inplane wave vector ${\mathbf k}_{in}$ as $\Gamma=4\pi^2 (dW/d\Omega)/(Sk^2 \hbar \omega_0\cos\theta)$. This justifies a purely classical treatment of luminescence by the exciton Bose condensate. 

First let us consider the limit of the light wave length long compared to the cloud radius $R$ [${\mathbf k}_{in} {\mathbf r} \to 0$ in equation~(\ref{flux})]. Then the distribution of the emitted power over the solid angle sphere is smooth being $\propto \sin^2 \theta$ for the dipole moment normal to the plane ($d_x=d_y=0$) and $\propto (1+ \cos^2 \theta)$ for the circularly polarized dipole moment normal in the plane ($d_z=0$, $d_y=\pm id_x$). But if the inplane dipole moment is linearly or elliptically polarized the axial symmetry of the distribution is broken, and the emitted power depends on the azimuthal angle $\phi$. The total emitted power does not depend on orientation of the dipole moment:
\begin{eqnarray}
W=\int_S{dW(\theta,\phi)\over d\Omega} d\Omega ={ck^4\over 3}{|{\mathbf d}|^2\over S} \left|\int_S \psi({\mathbf r})  \,d_2{\mathbf r}\right |^2.
                         \end{eqnarray}
In the vortex-free state, where the phase $\phi$ of $\psi({\mathbf r}) $ is constant, the power $W=ck^4|{\mathbf d}|^2N/3$ is proportional to the total number of Bose condensed excitons. However, if a vortex appear in the cloud center,  emission becomes impossible because of destructive phase interference  in the condensate ($\int \psi({\mathbf r})  \,d_2{\mathbf r}=0$). This is a strong effect, which can be used for experimental identification of Bose condensation  and vortex penetration.

Keeling {\em et al.}\cite{Levitov} have suggested to use the angular photoluminescence distribution for diagnostic of Bose condensation and vorticity in the opposite limit of the short wavelength, when there is a strong interference from the factor $e^{i{\mathbf k}_{in} {\mathbf r}}$ in equation~(\ref{flux}). Keeling {\em et al.}\cite{Levitov} considered the case, which corresponds to a rotating inplane dipole moment ($d_y=\pm i d_x$), when the interference leads to focusing of the radiation at the direction normal to the plane (the axis $z$, $\theta =0$). On the other hand, if the dipole moment oscillates along the axis $z$, the radiation flux vanishes at $\theta =0$ since the dipole does not radiate along its direction\cite{Jack}. Nevertheless focusing must take place in this case also. However, the  interference maximum is located not at $\theta=0$ but  at small  ring with radius $\theta \sim 1/kR$. If both the normal and the inplane dipole momenta are essential one may expect two maxima: at the center ($\theta=0$) and on the ring  ($\theta\neq 0$).

\begin{figure}[!t]
\begin{center}
   \leavevmode
  \includegraphics[width=0.6\linewidth]
  {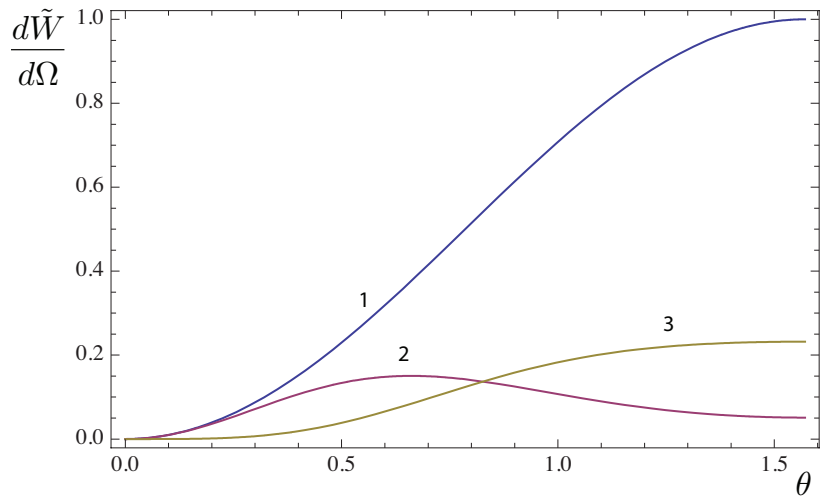}
  \caption{Angular dependence of the luminescence power for the axial dipole $d_z$. {\em 1} - the vortex-free state in the long wavelength limit $kR \to 0$. {\em 2} -  the vortex-free state, $kR=3$. {\em 3} -  the central-vortex  state, $kR=3$. The luminescence in the central-vortex  state at $kR \to 0$ vanishes. The dimensionless power on the vertical axis is $\tilde W =8\pi W/ck^4 |d_z|^2 N$.}
 \label{fig3}
 \end{center}
\end{figure}

In the short wavelength limit $kR \to \infty$ the effect of vortices is present  also, but the central vortex is not able to fully suppress radiation. The vortex effect can be evaluated with comparison of the coherence factor in equation~(\ref{flux}) for the vortex-free state, 
\begin{eqnarray}
{1\over S}\int_S \psi({\mathbf r}) e^{i{\mathbf k}_{in} {\mathbf r}} \,d_2{\mathbf r}={2 \sqrt{n_2}\over R^2} \int_0^R r\,dr\int _0^{2\pi} {d\varphi\over 2\pi} e^{ik_{in} r\cos \varphi}={2 \sqrt{n_2}\over k_{in} R } J_1(k_{in} R), 
     \end{eqnarray}
and for the state with a vortex in the center:
\begin{eqnarray}
{1\over S}\int_S \psi({\mathbf r}) e^{i{\mathbf k}_{in} {\mathbf r}} \,d_2{\mathbf r}={2 \sqrt{n_2}\over R^2} \int_0^R r\,dr\int _0^{2\pi} {d\varphi\over 2\pi} e^{i(\varphi+ k_{in} r\cos \varphi)}
={i \sqrt{n_2} k_{in} R\over 3 } \,_1F_2\left(\left\{3\over 2\right\}, \left\{2, {5\over 2}\right\}, -{k_{in}^2 R^2\over 4}\right),
     \end{eqnarray}
where $_1F_2\left(\left\{a\right\}, \left\{b, c\right\}, z\right)$ is the hypergeometric function and spatial variation of the  exciton density $n_2 =|\psi|^2$ was neglected. The effect of interference on the luminescence by the axial dipole $d_z$ is demonstrated  in Fig.~\ref{fig3}.  

In the present Letter we considered the exciton cloud  at zero temperature. Finite temperatures weaken discussed effects but cannot eliminate them as far as the ratio of the  superfluid and the total  exciton densities is not too small.

Though the analysis was focused on the exciton Bose-condensate the gauge-field rotation is a general phenomenon relevant for other neutral Bose particles.  Sato and Packard\cite{Pack} have suggested a direct measurement of the Aharonov-Bohm shift in superfluid $^4$He with the Sagnac superfluid interferometer in a radial electric and an axial magnetic field. This experiment is also possible in the 
geometry similar to that in the present Letter: a helium layer (instead of the exciton cloud in Fig.~\ref{fig1}) is subject to an axial electric field, which polarizes it, and to  a radial magnetic field.
The gauge-field rotation of the exciton-polariton Bose condensate is also possible. Vortices in the exciton-polariton Bose condensate can be spontaneously created  by the combined effect of pumping and inhomogeneity\cite{Berl}, while the gauge-field rotation would provide another controlled method of their creation. Vortices  in the exciton-polariton Bose condensate have a larger core size, and Lagoudakis {\sl et al.}\cite{Vort}  have already reported on the observation of vortices by  detection of the reduced polariton liquid density inside vortex cores with the  size $\sim 1~\mu$m.

In summary, it was shown that a moderate radial inplane magnetic field is able to rotate a Bose condensate of excitons in a double quantum well due to the Aharonov-Bohm effect for electrically polarized neutral excitons. This leads to penetration of vortices into the exciton cloud. Vortices strongly affect the photoluminescence from the Bose condensed exciton cloud. This can be used for their experimental detection. The existence of vortices is a direct consequence and evidence of exciton Bose-condensation.

I thank Richard Packard who attracted my attention to the Aharonov-Bohm  effect for electrically polarized neutral particles. I strongly appreciate interesting discussions with Ronen Rappaport and Leonid Shvartsman. The work was supported by the grant of the Israel Academy of Sciences and Humanities.

\end{document}